\documentclass[12pt,a4paper]{article}
\usepackage[inner=2cm,outer=1cm]{geometry}
\usepackage{graphicx}
\usepackage{caption}
\usepackage{subcaption}
\usepackage{amsmath}
\usepackage{amsfonts}
\usepackage{amssymb}
\usepackage{natbib}
\usepackage{epstopdf, epsfig}
\newcommand{\xo}{\left(x_1\right)}
\newcommand{\xt}{\left(x_2\right)}
\newcommand{\tu}{\tilde{u}}
\newcommand{\mbx}{\mathbf{x}}
\newcommand{\mbr}{\mathbf{r}}
\newcommand{\ol}[1]{\overline{#1}}

\newcommand{\lr}[1]{\left({#1}\right)}
\newcommand{\lrs}[1]{\left[{#1}\right]}
\newcommand{\pder}[2]{\frac{\partial{#1}}{\partial {#2}}}
\newcommand{\spv}{\mathnormal{v}}

\title{ Detailed derivation of scale-Space Energy Density Transport Equation for Compressible Inhomogeneous Turbulent Flows}

\author{S. Arun\thanks{Laboratoire de M\'{e}canique des Fluides et d'Acoustique, \'{E}cully, 69134, France},
 \and A. Sameen\thanks{Department of Aerospace Engineering, IIT Madras,Chennai, 600036, India},
 \and Balaji  Srinivasan\thanks{Department of Mechanical Engineering, IIT Madras,Chennai, 600036, India},
 \and Sharath S. Girimaji\thanks{Department of Ocean Engineering, Texas A\&M University, College Station, TX, USA}}

\begin{document}

\maketitle

\begin{abstract}
Scale-space energy density function ($E(\mbx,\mbr)$) is defined as the derivative of the two-point velocity correlation $Q_{ii}(\mbx,\mbr)$: $E(\mbx,\mbr) = -\frac{1}{2}\partial Q_{ii}(\mbx,\mbr)/\partial r$, where $\mbx$ is the spatial coordinate of interest and $\mbr$ is the separation vector. 
The function $E$ describes the turbulent kinetic energy density of scale $r$ at a location $x$ and can be considered as the generalization of spectral energy density function concept to inhomogeneous flows. 
We derive the transport equation for the scale-space energy density function in  compressible flows to develop a better understanding of scale-to-scale energy transfer and the degree of non-locality of the energy interactions. 
Specifically, the effects of variable-density and dilatation on turbulence energy dynamics are identified. 
It is expected that these findings will yield deeper insight into compressibility effects leading to improved models at all levels of closure for mass flux, density-variance, pressure-dilatation, pressure-strain correlation and dilatational dissipation processes.
\end{abstract}

\section{Introduction}\label{sec:intro}

Fluid turbulence is a non-linear, non-local phenomenon exhibiting the characteristics of a chaotic system. Much of the complexity of turbulence arises from the intricacies of energy exchange among different scales of motion (non-linearity effects) and nearly intractable interactions between different regions of the flow (non-locality effects).  Statistically-steady, incompressible homogeneous turbulence represents the simplest manifestation of this complex flow phenomenon. In this canonical case, energy distribution and exchange among scales can be conveniently considered in spectral space. Further, two-point correlations are dependent only on separation vector and hence amenable to simpler analysis. In such flows, Kolmogorov hypotheses \citep{K41,K62} lead to an insightful statistical description of energy distribution and exchange among the different spectral scales of motion. K\'{a}rm\'{a}n-Howarth-Monin (KHM) equations \citep{KH1938, Monin1959} govern the elementary multi-point statistical behavior. 
However, real-world turbulent flows are nearly always inhomogeneous featuring exponential instabilities, coherent structures and other spatio-temporal features. Further, compressibility can profoundly change the energy distribution and inter-scale energy transfer features of turbulence. 
For these flows, there is a pressing need to develop the mathematical framework for describing scale-to-scale energy transfer and elementary multi-point interactions. Such a development will further our physical understanding of these non-canonical effects and lead to improved closure models for flows of practical interest. 

Over the last twenty years, useful progress has been made to incorporate inhomogeneity effects into the classical analyses. \cite{Danaila2001} propose modifications to the KHM equation to account for lateral diffusion and inhomogeneity in channel flows. \cite{Hill2002} derives an exact equation for the second-order structure function, which applies to inhomogeneous and anisotropic turbulent flows. This generalized KHM equation provides a reasonable basis for analysis of interscale energy flux in anisotropic conditions and may be utilized for deriving sub-grid scale closure models -- \cite{Casciola2003}. In practical inhomogeneous flows, generalized KHM equation indicate the likelihood of  significant departures from the canonical scale-space energy transfer behaviour. For example, \cite{Cimarelli2012} point to the possibility of inverse energy transfer from small to large scales in strongly anisotropic inhomogeneous turbulence fields. By filtering the generalized KHM equations, the authors propose a methodology to improve the sub-grid scale models. \cite{Gomes2015} employ the generalized KHM equation to analyze the experimental data from grid-generated turbulence and demonstrate the occurrence of inverse cascade in the streamwise direction and forward cascade in the transverse direction. The KHM-based analysis of \cite{Mollicone2018} shows that the advection of energy in the joint  location-scale space causes vortical structures in the shear layer to be advected or dissipated. Thus the generalized KHM equation and subsequent modifications have enhanced our understanding of interscale energy transfer in anisotropic and inhomogeneous flows.

The next step in the natural progression toward further analytical development is to develop a physical space equivalent of spectral energy density function that is applicable to inhomogeneous turbulence fields.  Clearly, such a function must be positive semi-definite, and the integral of such a density function over the entire range of scales at a given location must yield the local turbulent kinetic energy. The trace of the second-order structure function is amenable to the interpretation that it represents (twice) the energy contained in all scales of size smaller than $r$. \cite{Davidson2005} proposes that the derivative of the second-order structure function can be considered as the physical scale-space energy density function as it possesses all of the requisite characteristics. For incompressible inhomogeneous flows, \cite{Hamba2015,Hamba2018} derives a transport equation for scale-space energy density function by taking an appropriate derivative of the two-point correlation function. The author also demonstrates that filtering the transport equation can directly lead to insight and subgrid closure modeling guidance for important turbulence processes such a pressure-strain correlation, transport and dissipation in incompressible turbulent flows.

The progress in including compressibility effects into KHM and scale-space energy analysis has been more limited. \cite{Clark1995} develop two-point velocity correlation function transport equation for variable-density turbulence. \cite{Lai2018} derive the variable-density KHM equations and demonstrate the effects of non-uniform density on the interscale energy transfer.
While the above studies address variable-density physics, compressibility effects arising from high Mach numbers are not considered. The emergence of dilatational velocity field in high-speed compressible flows leads to fundamental changes in energy interactions that are not encountered in low-speed variable density flows. \cite{Mittal2019} propose a basic framework for examining turbulent kinetic and internal energy interactions in such flows. \cite{Praturi2019} derive the spectral energy density equations for homogeneous compressible flows. However, the fundamental changes in turbulence dynamics (non-locality of different processes and energy interactions) arising from inhomogeneous dilatational velocity field remains to be analyzed and investigated.

Thus, the objective of this work is to derive the transport equation for the scale-space energy density function in inhomogeneous variable-density and compressible flows.

\section{Scale-space energy density}\label{sec:energy}
\subsection{Energy density definition}
Two-point velocity correlation function forms the basis of scale-space energy density function \citep{Hamba2015}.
In compressible or variable density flows, the definition of two-point correlation must be generalized to include the effects of density variation in space \citep{Clark1995}.
Thus we use the following definition of two-point correlation, 
\begin{equation}
Q_{ij}(\mbx,\mbr)=\frac{1}{2}\overline{\lrs{\rho(\mbx)+\rho(\mbx+\mbr)}u_i''(\mbx)u_j''(\mbx+\mbr)},
\label{eq:2pt}
\end{equation}
where $u_i''$ is the fluctuation from Favre averaged velocity $\left(\tilde{u}=\overline{\rho u}/\bar{\rho}\right)$.
For $r=0$, $Q_{ij}$ reduces to the Reynolds stress $R_{ij}(\mbx)$.
Then, the trace of the tensor yields the volume-averaged turbulent kinetic energy, $K(\mbx)=R_{ii}(\mbx)/2=Q_{ii}(\mbx,0)/2$.
The correlations decay to zero in a fully developed turbulent flow as the separation distance becomes large, $Q_{ij}(\mbx,\infty)=0$.
Then, the energy density in scale-space is identified as
\begin{equation}
E(\mbx,\mbr)=-\frac{1}{2}\frac{\partial}{\partial r}Q_{ii}(\mbx,\mbr).
\label{eq:energy}
\end{equation}
This definition is consistent with that proposed by \cite{Hamba2015} for incompressible turbulent flows, except for the inclusion of fluid density in $Q_{ii}$.

Turbulent kinetic energy and the Reynolds stress are given by $K=\frac{1}{2}\overline{\rho u_i''u_i''}$ and $R_{ij}=\overline{\rho u_i''u_j''}$ respectively.
It must be noted that the turbulent kinetic energy defined here is a volume-averaged quantity, not a mass-averaged one as in the case of incompressible turbulence.
In inhomogeneous turbulent flows, these quantities are functions of the spatial coordinates.
Therefore,at a given time, both $K$ and $R_{ij}$ are functions of $\mbx$ in general. 
The turbulent kinetic energy and energy density in scale space are related by,
\begin{equation}
K(\mbx)=\int_0^\infty E(\mbx,\mbr)\:\mathrm{d}r,
\label{eq:tke}
\end{equation}
implying that the turbulent kinetic energy at any spatial location is the integral of the energy density over all scales at that location. 
In other words, $E(r)dr$ is the amount of energy residing in eddies of size $r$ to $r+dr$.
In homogeneous turbulence, where a Fourier representation is possible, there is a similar relation - $E(k)dk$ is the energy content of wavenumbers in a spherical shell of 
radius $k$. 
Therefore, this new energy density definition is equivalent to the energy spectrum in Fourier space.
However, unlike the latter, the applicability of $E(\mbx,\mbr)$ is not constrained by the homogeneity requirement.
This makes the energy density a more appropriate choice for inhomogeneous flows in nature and engineering.

\subsection{Transport equation for $Q_{ij}(\mbx,\mbr)$}
Transport equations for quantities such as turbulent kinetic energy, second-order structure function and energy spectrum provide invaluable insight into
the turbulence dynamics. 
Hence, it is of much value to develop a transport equation for the newly defined energy density to gain an in-depth understanding of the turbulent dynamics across different
scales of motion.
By definition \eqref{eq:energy}, $E(\mbx,\mbr)$ is the derivative of the two-point correlation. 
Therefore, an exact equation for the energy density function starts with that for $Q_{ij}(\mbx,\mbr)$.
A detailed derivation of the two-point correlation transport equation (\cite{Clark1995}) is derived from the compressible Navier-Stokes equations in
the Appendix. For the sake of completeness, the key steps are outlined here.

The pressure, density and velocity fields in compressible turbulent flows are decomposed to a mean and fluctuating part.
Reynolds averaging (overbar, $\bar{.}$) is employed for pressure and density fields, whereas a density-weighted or Favre averaging (tilde, $\tilde{.}$) is used for the velocity field.
\begin{equation}
p=\bar{p}+p',\ \rho=\bar{\rho}+\rho',\ u=\tilde{u}+u''.
\end{equation}
Fluctuations from Reynolds and Favre averages are denoted by $\left(\:'\right)$ and $\left(\:''\right)$ respectively.
It is possible to express the fluctuating velocity field equation in two forms - one for $u''$ and another for $\rho u''$.
\begin{equation}
\frac{\partial u''_i}{\partial t}+u''_k\pder{\lr{\tilde{u}_i+u''_i}}{x_k}+\tilde{u_k}\pder{u''_i}{x_k}=\lr{\bar{\spv}-\frac{1}{\bar{\rho}}}\pder{\bar{\sigma}_{ik}}{x_k}
+\spv'\pder{\bar{\sigma}_{ik}}{x_k}+\spv\pder{\sigma'_{ik}}{x_k}+\frac{1}{\bar{\rho}}\pder{\bar{\rho}R_{ik}}{x_k},
\label{fluc_v}
\end{equation}
\begin{equation}
\frac{\partial \rho u''_i}{\partial t}+\pder{ }{x_k}\lr{\rho u''_i\lrs{\tilde{u}_k+u''_k}}+\rho u''_k\pder{\tilde{u}_k}{x_k}=\pder{\sigma'_{ik}}{x_k}
+\lr{1-\frac{\rho}{\bar{\rho}}}\pder{\bar{\sigma}_{ik}}{x_k}+\frac{\rho}{\bar{\rho}}\pder{\bar{\rho}R_{ik}}{x_k}.
\label{fluc_rv}
\end{equation}
Here $\sigma_{ij}=p\delta_{ij}+\tau_{ij}$, where $\tau_{ij}$ is the viscous stress tensor and $\spv$ is the specific volume.

For compressible and variable density flows, the correlation between points $\mbx$ and $\mbx'$ is the average of 
$\lrs{\rho\lr{\mbx}+\rho\lr{\mbx'}}u_i''\lr{\mbx}u_j''\lr{\mbx'}$.
The equation for this quantity is obtained from equations \ref{fluc_v} and \ref{fluc_rv} after suitable manipulations:
\begin{align}
\nonumber \pder{}{t}\overline{\lrs{\rho\lr{\mbx}+\rho\lr{\mbx'}}u_i''\lr{\mbx}u_j''\lr{\mbx'}} & =\overline{ \rho\lr{\mbx'}u_j''\lr{\mbx'}\pder{ u_i''}{t}\lr{\mbx}}+\overline{u_i''\lr{\mbx}\pder{}{t}\lr{\rho\lr{\mbx'}u_j''\lr{\mbx'}}} \\
 & +\overline{\rho\lr{\mbx}u_i''\lr{\mbx}\pder{u_j''}{t}\lr{\mbx'}}+\overline{u_j''\lr{\mbx'}\pder{}{t}\lr{\rho\lr{\mbx}u_j''\lr{\mbx}}}.
\end{align}
Similar operations are performed for all the terms in equations \ref{fluc_v} and \ref{fluc_rv}.
Then, setting $\mbx'=\mbx+\mbr$, the spatial derivatives at $\mbx$ and $\mbx'$ transform to,
\begin{flalign}
    \left.\pder{}{x_k}\right\vert_\mbx=\pder{}{x_k}-\pder{}{r_k}; \qquad 
    \left.\pder{}{x_k}\right\vert_{\mbx'}=\pder{}{r_k}.
\end{flalign}
This transformation  brings in derivatives with respect to the separation vector which are crucial to achieving the objective of representing the dynamics in scale-space.
Similar transformations of the independent variables are employed by \cite{Hill2002} to obtain exact equations of second-order structure-function in incompressible flows.
Following this, the transport equation for two-point correlation is given by (see Appendix and \citep{Clark1995} for further details),
\begin{flalign}
&\pder{Q_{ij}(\mbx,\mbr)}{t}+\pder{Q_{ij}(\mbx,\mbr)\tilde{u}_k(\mbx)}{x_{k}}= 
\underbrace{-Q_{kj}(\mbx,\mbr)\pder{\tilde{u}_i(\mbx)}{x_{k}}-Q_{ik}(\mbx,\mbr)\pder{\tilde{u}_j(\mbx+\mbr)}{x_{k}}}_{\mathcal{P}_{ij}} 
\nonumber \\
&\underbrace{-\pder{T_{ijk}(\mbx,\mbr)}{x_{k}}}_{\mathcal{D}^u_{ij}}+\underbrace{\frac{1}{2}\lrs{\Phi_{ij}^p(\mbx,\mbx+\mbr)+\Phi_{ji}^p(\mbx+\mbr,\mbx)}}_{\mathcal{D}_{ij}^p}+\underbrace{\frac{1}{2}\lrs{\Phi_{ij}^\nu(\mbx,\mbx+\mbr)+\Phi_{ji}^\nu(\mbx+\mbr,\mbx)}}_{\mathcal{D}_{ij}^\nu} \nonumber \\
&+\underbrace{\frac{1}{2}\lrs{\Psi_{ij}^p(\mbx,\mbx+\mbr)+\Psi_{ji}^p(\mbx+\mbr,\mbx)}}_{\Pi_{ij}}+\underbrace{\frac{1}{2}\lrs{\Psi_{ij}^\nu(\mbx,\mbx+\mbr)+\Psi_{ji}^\nu(\mbx+\mbr,\mbx)}}_{\epsilon_{ij}} \nonumber \\
&\underbrace{+\frac{1}{2}Q_{ij}(\mbx,\mbr)\lrs{\pder{\tilde{u}_k(\mbx)}{x_{k}}+\pder{\tilde{u}_k(\mbx+\mbr)}{x_{k}}}}_{\chi_{ij}} 
\underbrace{+\frac{1}{2}Q^{(-)}_{ij}(\mbx,\mbr)\lrs{\pder{\tilde{u}_k(\mbx+\mbr)}{x_{k}}-\pder{\tilde{u}_k(\mbx)}{x_{k}}}}_{\chi_{ij}^-} \nonumber \\
&\underbrace{+\overline{\rho(\mbx+\mbr)u''_i(\mbx)u''_j(\mbx+\mbr)\pder{u''_k(\mbx)}{x_k}}+\overline{\rho(\mbx)u''_i(\mbx)u''_j(\mbx+\mbr)\pder{u''_k(\mbx+\mbr)}{x_k}}}_{\delta_{ij}} \nonumber \\
&\underbrace{+\frac{1}{2}\lr{a_i(\mbx,\mbx+\mbr)\pder{\bar{\sigma}_{jk}(\mbx+\mbr)}{x_k}+a_j(\mbx+\mbr,\mbx)\pder{\bar{\sigma}_{ik}(\mbx+\mbr)}{x_k}}}_{\Sigma_{ij}} 
\underbrace{+\frac{1}{2}\lrs{c_{ij}(\mbx,\mbx+\mbr)+c_{ji}(\mbx+\mbr,\mbx)}}_{\mathcal{R}_{ij}} \nonumber \\
&\underbrace{-\pder{}{r_{k}}\lrs{\frac{1}{2}\overline{\lr{\rho(\mbx)+\rho(\mbx+\mbr)}u_i''(\mbx)u_j''(\mbx+\mbr)(u_k''(\mbx+\mbr)-u_k''(\mbx))} 
+\lr{\tilde{u}_k(\mbx+\mbr)-\tilde{u}_k(\mbx)}Q_{ij}(\mbx,\mbr)}}_{\mathcal{T}_{ij}},
\label{eq:2pt_transport}
\end{flalign}
where the various correlations are,
\begin{eqnarray}
Q_{ij}^{(-)}(\mbx,\mbr) & =& \frac{1}{2}\overline{\lrs{\rho(\mbx)-\rho(\mbx+\mbr)}u_i''(\mbx)u_j''(\mbx+\mbr)}\:,\\
T_{ijk}(\mbx,\mbr) &=& \frac{1}{2}\overline{\lr{\rho(\mbx)+\rho(\mbx+\mbr)}u_i''(\mbx)u_j''(\mbx+\mbr)u_k''(\mbx)}, \\
\Psi_{ij}^p(\mbx,\mbx+\mbr)&=&\overline{\lrs{1+\frac{\rho(\mbx)}{\rho(\mbx+\mbr)}}p'(\mbx+\mbr)\pder{u_i''(\mbx)}{x_{j}}}, \\
\Psi_{ij}^\nu(\mbx,\mbx+\mbr)&=&-\overline{\lrs{1+\frac{\rho(\mbx)}{\rho(\mbx+\mbr)}}\tau'_{jk}(\mbx+\mbr)\pder{u_i''(\mbx)}{x_{k}}}, \\
\Phi_{ij}^p(\mbx,\mbx+\mbr) &=& -\overline{\lrs{1+\frac{\rho(\mbx)}{\rho(\mbx+\mbr)}}\pder{u''_i(\mbx)p'(\mbx+\mbr)}{x_k}\delta_{jk}} \\
\Phi_{ij}^\nu(\mbx,\mbx+\mbr) &=& \overline{\lrs{1+\frac{\rho(\mbx)}{\rho(\mbx+\mbr)}}\pder{u''_i(\mbx)\tau'_{jk}(\mbx+\mbr)}{x_k}} \\
a_i(\mbx,\mbx+\mbr) & =& -\overline{\rho(\mbx)u_i''(\mbx)\spv(\mbx+\mbr)}-\overline{u_i''(\mbx)}+\frac{\overline{\rho(\mbx+\mbr)u_i''(\mbx)}}{\bar{\rho}(\mbx+\mbr)},\\
c_{ij}(\mbx,\mbx+\mbr) & =& \frac{\overline{\rho(\mbx+\mbr)u_i''(\mbx)}}{\bar{\rho}(\mbx+\mbr)}\pder{R_{jk}(\mbx+\mbr)}{x_k}\:.
\end{eqnarray}

The term $\mathcal{P}_{ij}$ represents the production of two-point correlation and $\mathcal{T}_{ij}$ is the turbulent diffusion of $Q_{ij}(\mbx,\mbr)$.  
The effect of fluctuating viscous stress and pressure are accounted in the term $\Psi_{ij}$. 
The other mechanisms that influence the two-point correlation are mean flow dilatation $\lr{\chi_{ij}}$, inhomogeneity of mean field dilatation $\lr{\chi_{ij}^-}$,
fluctuating field dilatation $\lr{\delta_{ij}}$, mean stress $\lr{\Sigma_{ij}}$ and turbulent stress $\lr{\mathcal{R}_{ij}}$.
The last line, which are gradients in scale space, is the transport across the scales - one part of which is due to the mean velocity and the other due to the fluctuating field.
These interscale transfer terms originate from the non-linear part of the Navier-Stokes equation.

\subsection{Transport equation for $E(\mbx,\mbr)$}
The scale-space energy density function $E(\mbx,\mbr)$ is obtained by differentiating the derivative of the trace of the two-point correlation equation \ref{eq:2pt_transport}  
with respect to $r$:
\begin{equation}
	\frac{DE(\mbx,\mbr)}{Dt}=\mathcal{P}_r+\mathcal{D}^u_r+\mathcal{D}^p_r+\mathcal{D}^\nu_r-\epsilon_r+\mathcal{T}_r+\Pi_r+\chi_r+\delta_r+\Sigma_r+\mathcal{R}_r,
\label{eq:transport}
\end{equation}
where $\frac{D}{Dt}=\pder{}{t}+\tilde{u}_k(\mbx)\pder{}{x_k}$.
The right hand side of equation \ref{eq:transport} describes the various physical mechanisms influencing the distribution of turbulent kinetic energy across scales at any
given location in physical space.
Each of these terms is discussed in detail below.

The first of these is the production of energy density,
\begin{equation}
\mathcal{P}_r=-\frac{1}{2}\pder{\mathcal{P}_{ii}}{r}=\frac{1}{2}\pder{Q_{ki}(\mbx,\mbr)}{r}\lr{\pder{\tilde{u}_i(\mbx)}{x_k}+\pder{\tilde{u}_k(\mbx)}{x_i}}+\frac{1}{2}\pder{}{r}\lr{Q_{ik}(\mbx,\mbr)\pder{\lr{\tilde{u}_i(\mbx+\mbr)-\tilde{u}_i(\mbx)}}{x_k}},
\label{production}
\end{equation}
which is equivalent to the turbulent kinetic energy production by mean shear. 
The production of energy density describes how extraction of energy from mean flow by turbulence varies across different scales of motion.
The first part of $\mathcal{P}_r$ is due to the mean shear at the physical location $\mbx$, whereas the second part accounts for spatial difference in mean shear which implies
that statistical inhomogeneity alters how turbulent energy is produced at different scales.
There are no explicit compressibility effects associated with production.

The transport terms - turbulent, pressure and viscous respectively are considered next:
\begin{equation}
\mathcal{D}^u_r=-\frac{1}{2}\pder{\mathcal{D}_{ii}^u}{r}=\frac{1}{4}\pder{}{x_k}\lr{\pder{\overline{\lr{\rho(\mbx)+\rho(\mbx+\mbr)}u_i''(\mbx)u_i''(\mbx+\mbr)u_k''(\mbx)}}{r}},
\label{ttransport}
\end{equation}
\begin{align}
\mathcal{D}_r^p=-\frac{1}{2}\pder{\mathcal{D}_{ii}^p}{r}=\frac{1}{4}\pder{}{r}\Bigg(&\overline{\lrs{1+\frac{\rho(\mbx)}{\rho(\mbx+\mbr)}}\pder{u_k''(\mbx)p'(\mbx+\mbr)}{x_k}} \nonumber \\
&+ \overline{\lrs{1+\frac{\rho(\mbx+\mbr)}{\rho(\mbx)}}\pder{u_k''(\mbx+\mbr)p'(\mbx)}{x_k}}\Bigg),
\label{ptransport}
\end{align}
\begin{align}
\mathcal{D}_r^\nu=-\frac{1}{2}\pder{\mathcal{D}_{ii}^\nu}{r}=-\frac{1}{4}\pder{}{r}\Bigg(&\overline{\lrs{1+\frac{\rho(\mbx)}{\rho(\mbx+\mbr)}}\pder{u_i''(\mbx)\tau_{ik}'(\mbx+\mbr)}{x_k}}\nonumber \\
			&+\overline{\lrs{1+\frac{\rho(\mbx+\mbr)}{\rho(\mbx)}}\pder{u_i''(\mbx+\mbr)\tau_{ik}'(\mbx)}{x_k}}\Bigg).
\label{vtransport}
\end{align}
All these transport terms have equivalent counter parts in the single-point statistics budget and reduces to those respective terms on integration in scale space.
These terms provide information about the scales at which these processes occur.
All these diffusion contributions vanish in homogeneous turbulent flow. In variable-density turbulence, these terms may play a significant role.

The viscous dissipation of scale-space energy density is given by,
\begin{align}
\epsilon_r=\frac{1}{2}\pder{\epsilon_{ii}}{r}=-\frac{1}{4}\pder{}{r}\Bigg(&\overline{\lrs{1+\frac{\rho(\mbx)}{\rho(\mbx+\mbr)}}\tau_{ik}'(\mbx+\mbr)\pder{u_i''(\mbx)}{x_k}} \nonumber \\
&+\overline{\lrs{1+\frac{\rho(\mbx+\mbr)}{\rho(\mbx)}}\tau_{ik}'(\mbx)\pder{u_i''(\mbx+\mbr)}{x_k}}\Bigg).
\label{dissipation}
\end{align}
This term represents the dissipation of turbulent kinetic energy at a given scale. 
Dissipation is a characteristic feature of turbulent flows irrespective of whether the flow is compressible or incompressible, homogeneous or inhomogeneous.
Our existing knowledge of turbulence is that dissipation is a small scale phenomenon.

One of the important aspects of the turbulent scale dynamics is scale energy transfer by which energy is transferred from the energy producing large scales to the 
dissipative small scales.
The interscale transfer term in the above budget equation,
\begin{align}
\nonumber \mathcal{T}_r=-\frac{1}{2}\pder{\mathcal{T}_{ii}}{r}=\frac{1}{4}\pder{}{r}\pder{}{r_{k}}\bigg(\overline{\lrs{\rho(\mbx)+\rho(\mbx+\mbr)}u_i''(\mbx)u_i''(\mbx+\mbr)}[\tilde{u}_k(\mbx+\mbr)-\tilde{u}_k(\mbx)]\bigg) \\ 
+\frac{1}{4}\pder{}{r}\pder{}{r_{k}}\bigg(\overline{\lrs{\rho(\mbx)+\rho(\mbx+\mbr)}u_i''(\mbx)u_i''(\mbx+\mbr)[u_k''(\mbx+\mbr)-u_k''(\mbx)]}\bigg),
\label{ist}
\end{align}
quantifies the energy pathways across the scales.
The first part of equation \ref{ist} features energy transfer due to mean velocity differences, indicating that spatial inhomogeneity is key to interscale transfer mechanism.
The second part is due to differences in velocity fluctuations.
In KHM literature these are called linear and non-linear transfer, respectively \citep{Portela2017}.
\cite{Cimarelli2012} point out the inhomogeneity can give rise to inverse transfer of energy from small to large scales.
The linear transfer term accounts for these inhomegeneities.
The interscale transfer term originates from the non-linearity of the compressible Navier-Stokes equations, and being a wholly scale-space mechanism, 
is not accounted for in the turbulent kinetic energy budget.
The nonlinear interscale transfer, like dissipation, is relevant in homogeneous and inhomogeneous cases, but the linear contribution is only present in 
inhomogeneous flow field.
The interscale transfer can alternatively be written as the gradient in scale space of an energy flux, either as
\begin{equation}
    \mathcal{T}_r=-\pder{}{r}\Pi_E,\ \Pi_E=-\frac{1}{4}\pder{}{r_k}\left(\overline{\lrs{\rho(\mbx)+\rho(\mbx+\mbr)}u''_i(\mbx)u''_i(\mbx+\mbr)\lrs{u_k(\mbx+\mbr)-u_k(\mbx)}}\right)
    \label{scalarflux}
\end{equation}
or as,
\begin{align}
\mathcal{T}_r=-\nabla_r.F_E,\ F_E=-\frac{1}{4}\pder{}{r}\left(\overline{\lrs{\rho(\mbx)+\rho(\mbx+\mbr)}u''_i(\mbx)u''_i(\mbx+\mbr)\lrs{u_k(\mbx+\mbr)-u_k(\mbx)}}\right).
\label{vectorflux}
\end{align}
The scalar flux in equation \ref{scalarflux} represents energy flux for a given scale size $r$, whereas $F_E$ represents the flux in three-dimensional scale-space.
The latter simplifies analysis of direction of energy transfer in the scale-space, similar to those found in KHM literature \citep{Mollicone2018}.

Pressure-dilatation effect which is exclusive to compressible flows is given by,
\begin{align}
\Pi_r=-\frac{1}{2}\pder{\Pi_{ii}}{r}=-\frac{1}{4}\pder{}{r}\Bigg(&\overline{\lrs{1+\frac{\rho(\mbx)}{\rho(\mbx+\mbr)}}p'(\mbx+\mbr)\pder{u_i''(\mbx)}{x_i}} \nonumber \\
&+\overline{\lrs{1+\frac{\rho(\mbx+\mbr)}{\rho(\mbx)}}p'(\mbx)\pder{u_i''(\mbx+\mbr)}{x_i}}\Bigg),
\label{pdil}
\end{align}
and orginates from $\Pi_{ij}$ term in equation \ref{eq:2pt_transport}.
This term upon integration over the scale-space yields the pressure-strain correlation term in turbulent kinetic energy budget.
This is a compressible flow phenomenon which is relevant in both homogeneous and inhomogeneous turbulence.
The effects of pressure-strain correlation in compressible turbulent flows has been the subject of various studies over the years.
The $\Pi_r$ term in equation \ref{eq:transport} may be employed to analyse the scales at which redistribution of kinetic energy by pressure occurs and if there is any transfer across scales due to the effects of pressure.

The remaining terms in the transport equation stem fr0m the effects of compressibility.
The contribution from mean flow dilatation,
\begin{align}
\nonumber \chi_r=&-\frac{1}{2}\pder{}{r}\lr{\chi_{ii}^-+\chi_{ii}}-E(\mbx,\mbr)\pder{\tilde{u}_k}{x_k} \\
=&-\frac{1}{4}Q_{ii}(\mbx,\mbr)\pder{}{r}\lr{\pder{\tilde{u}_k(\mbx+\mbr)}{x_k}}+\frac{1}{2}E(\mbx,\mbr)\lrs{\pder{\tilde{u}_k(\mbx+\mbr)}{x_k}-\pder{\tilde{u}_k(\mbx)}{x_k}} \nonumber \\
&-\frac{1}{4}\pder{}{r}\lr{Q_{ii}^{(-)}(\mbx,\mbr)\lrs{\pder{\tilde{u}_k(\mbx+\mbr)}{x_k}-\pder{\tilde{u}_k(\mbx)}{x_k}}}, 
\label{mdil}
\end{align}
accounts for inhomogeneity of mean field dilatation and combines the effects of $\chi_Q$ and $\chi_Q^-$.
The effect vanishes in homogeneous turbulent flows.
Integral of $\chi_r$ over the scale space is zero which implies that it is a scale-space phenomenon.
Therefore, dilatation of mean velocity field plays a role in the scale-space dynamics in compressible turbulent flows.

The fluctuating field dilatation effects on the scale-space energy density is given by,
\begin{align}
\nonumber \delta_r=-\frac{1}{2}\pder{\delta_{ii}}{r}
=-\frac{1}{2}\pder{}{r}\Bigg(&\overline{\rho(\mbx+\mbr)u''_i(\mbx)u''_j(\mbx+\mbr)\pder{u''_k(\mbx)}{x_k}}\\
 &+\overline{\rho(\mbx)u''_i(\mbx)u''_j(\mbx+\mbr)\pder{u''_k(\mbx+\mbr)}{r_k}}\Bigg)
\label{fdil}
\end{align}
also originate from the non-linear terms in Navier-Stokes equations.

The contribution from mean stress effects,
\begin{equation}
\Sigma_r=-\frac{1}{2}\pder{\Sigma_{ii}}{r}=-\frac{1}{4}\pder{}{r}\lr{a_i(\mbx,\mbx+\mbr)\pder{\bar{\sigma}_{ik}(\mbx+\mbr)}{x_{k}}+a_i(\mbx+\mbr,\mbx)\pder{\bar{\sigma}_{ik}(\mbx)}{x_{k}}},
\label{mstress}
\end{equation}
corresponds to the mass-flux contributions to turbulent kinetic energy budget as shown in \cite{Gatski2009}.
Even though mass-flux effects on turbulent kinetic energy budget are found to be very negligible in most cases, the scale-space representation makes it possible to analyze if it plays a role in energy transfer in the scale-space.

The turbulent stress term, 
\begin{equation}
\mathcal{R}_r=-\frac{1}{2}\pder{\mathcal{R}_{ii}}{r}=-\frac{1}{4}\pder{}{r}\lr{c_i(\mbx,\mbx+\mbr)\pder{R_{ik}(\mbx+\mbr)}{x_{k}}+c_i(\mbx+\mbr,\mbx)\pder{R_{ik}(\mbx)}{x_{k}}}
\label{tstress}
\end{equation} 
vanishes on integration over scale-space, which implies that it is a scale-space phenomenon.
This term indicates that spatial variation of turbulent stress, coupled with density fluctuations, has an impact on the turbulent flow dynamics at different scales of motion.
Both the mean and turbulent stress effects are due to non-uniform density field which is a feature of compressible flows.
Similar terms, therefore, appear in the variable-density generalisation of KHM equation by \cite{Lai2018}.

\textbf{Homogeneous turbulence}: all terms, except the dilatation contributions, viscous dissipation  
and energy transfer in scale space due to fluctuating velocity, vanish. 
\begin{equation}
    \pder{E}{t}=-\epsilon_r+\Pi_r+\mathcal{T}_r+\delta_r
    \label{eq:hit}
\end{equation}
The terms on the right-hand side results in energy transfer across scales as well as to internal energy.
Apart from the viscous dissipation, these phenomena are due to the non-linear inertial effects and the action of pressure, for which \cite{Praturi2019} provide an equation in the spectral space.
Equation \ref{eq:hit} is a much simplified form of the transport equation \ref{eq:transport}, which implies that
inhomogeneity introduces a lot more dynamics into the system from the perspective of energy density in scale space.

\textbf{Incompressible limit}: the effects of dilatation and variable density are absent. The transport equation for energy density function reduces to
\begin{equation}
\frac{DE}{Dt}=\mathcal{P}_r+\mathcal{D}^u_r+\mathcal{D}^p_r+\mathcal{D}^\nu_r-\epsilon_r+\mathcal{T}_r,
\end{equation}
which is same as that given by \cite{Hamba2015}.
If the flow is homogenous, the equation further simplifies to
\begin{equation}
    \pder{E}{t}=-\epsilon_r+\mathcal{T}_r.
\end{equation}
In comparison, the compressible variant of the equation has dilatation effects which influences energy transfer across scales as well as that between kinetic and internal energies.
\section{Summary and conclusion}\label{sec:conclusion}
The scale-space energy density function $E(\mbx,\mbr)$ describes the energy intensity in different scales ($\mbr$) of turbulent motion at various locations ($\mbx$) in the flow field. The function $E(\mbx,\mbr)$ not only extends the concept of spectral energy density distribution to inhomogeneous flows, but it also describes detailed energy interactions between any two locations in the flow field. In this work, the transport equations for $E(\mbx,\mbr)$ in variable-density and compressible flows is developed by differentiating the  trace of density-weighted two-point velocity correlation tensor $Q(\mbx,\mbr)$ \citep{Clark1995} with respect to separation distance. In the limit of vanishing dilatation and uniform density, the new transport equation simplifies to that derived in \cite{Hamba2015} for incompressible flows. In homogeneous turbulence, the present equation is consistent with the spectral energy density equation of \cite{Praturi2019}. 

The integral of $E(\mbx,\mbr)$ over the entire range of scales yields the turbulent kinetic energy budget equation at a given spatial location. The derivation procedure provides valuable insight into the scale-distribution and degree of non-locality of compressibility effects on energy production, dissipation and turbulent transport. Such an equation will serve as the foundation for examining the influence of shocks, chemical reaction (combustion), compression/expansion waves and dilatational flow structures on scale-to-scale energy transfer and non-locality effects in inhomogeneous high Mach number turbulent flows. Filtering the scale-space energy density function along the lines of \cite{Hamba2018} can lead to the development of advanced cut-off dependent closure models for practical flow computations

\appendix
\section*{Appendix: Derivation of the exact transport equation}
\noindent The energy density function is defined in terms of the two-point velocity correlation and therefore, a transport equation for energy density is derived from that for two-point correlation. The derivation starts from the compressible Navier-Stokes equation \citep{Clark2016}.
\begin{equation}
\pder{\rho u_i}{t}+\pder{\rho u_iu_j}{x_j}=\pder{\sigma_{ij}}{x_j},
\label{nse}
\end{equation}
where $\sigma_{ij}=-p\delta_{ij}+\tau_{ij}$ and 
\begin{equation}
\tau_{ij}=\mu\lr{\pder{u_i}{x_j}+\pder{u_j}{x_i}-\frac{2}{3}\pder{u_k}{x_k}}.
\end{equation}
The mass conservation equation is,
\begin{equation}
\pder{\rho}{t}+\pder{\rho u_k}{x_k}=0.
\label{ce}
\end{equation}
From \eqref{nse} and \eqref{ce}, exact equation for instantaneous velocity is given by,
\begin{equation}
\pder{u_i}{t}+u_j\pder{u_i}{x_j}=\frac{1}{\rho}\pder{\sigma_{ij}}{x_j}.
\label{instu}
\end{equation}
Applying Favre-averaging to the momentum and mass conservation equations,
\begin{align}
\pder{\bar{\rho}\tu_i}{t}+\pder{\bar{\rho}\tu_i\tu_j}{x_j}+\pder{R_{ij}}{x_j}=&\pder{\bar{\sigma}_{ij}}{x_j}, \\
\pder{\bar{\rho}}{t}+\pder{\bar{\rho}\tu_k}{x_k}=&0,
\end{align}
where $R_{ij}=\ol{\rho u''_iu''_j}$ is the turbulent stress. 
From these an exact equation for $\tu_i$ is obtained as,
\begin{equation}
\pder{\tu_i}{t}+\tu_j\pder{\tu_i}{x_j}+\frac{1}{\bar{\rho}}\pder{R_{ij}}{x_j}=\frac{1}{\bar{\rho}}\pder{\bar{\sigma}_{ij}}{x_j}.
\label{meanu}
\end{equation}
Then the equation for fluctuating velocity is obtained by subtracting \eqref{meanu} from \eqref{instu},
\begin{equation}
\pder{u''_i}{t}+u_j\pder{u''_i}{x_j}+u''_j\pder{\tu_i}{x_j}-\frac{1}{\bar{\rho}}\pder{R_{ij}}{x_j} = \frac{1}{\rho}\pder{\sigma'_{ij}}{x_j}+\lr{\spv-\frac{1}{\bar{\rho}}}\pder{\bar{\sigma}_{ij}}{x_j},
\end{equation}
where $\spv=1/\rho$ is the specific volume. However, $1/\bar{\rho}\neq\bar{\spv}$.
This equation can be rewritten in the conservative form, by making use of \eqref{ce}, as
\begin{equation}
\pder{\rho u''_i}{t}+\pder{\rho u''_iu_j}{x_j}+\rho u''_j\pder{\tu_i}{x_j}-\frac{\rho}{\bar{\rho}}\pder{R_{ij}}{x_j} = \pder{\sigma'_{ij}}{x_j}+\lr{1-\frac{\rho}{\bar{\rho}}}\pder{\bar{\sigma}_{ij}}{x_j}.
\end{equation}

\subsubsection*{Two-point correlation transport equation}
To derive the two-point correlation transport equation we consider two points $x_1$ and $x_2$. 
Fluctuating velocity equation at $x_1$ and $x_2$ in their primitve and conservative forms are as follows.
\begin{subequations}
\begin{align}
\nonumber \pder{u''_i\xo}{t}+u_k\xo\pder{u''_i\xo}{x_k}+u''_k\xo\pder{\tu_i\xo}{x_k}-\frac{1}{\bar{\rho}\xo}\pder{R_{ik}\xo}{x_k} =  &\\
\frac{1}{\rho\xo}\pder{\sigma'_{ik}\xo}{x_k}+\lr{\spv\xo-\frac{1}{\bar{\rho}\xo}}\pder{\bar{\sigma}_{ik}\xo}{x_k},& \label{primi1}\\
\nonumber \pder{\rho\xo u''_i\xo}{t}+\pder{\rho\xo u''_i\xo u_k\xo}{x_k}+\rho\xo u''_k\xo\pder{\tu_i\xo}{x_k}-\frac{\rho\xo}{\bar{\rho}\xo}\pder{R_{ik}\xo}{x_k} = & \\
\pder{\sigma'_{ik}\xo}{x_k}+\lr{1-\frac{\rho\xo}{\bar{\rho}\xo}}\pder{\bar{\sigma}_{ik}\xo}{x_k}, &\label{consi1}\\
\nonumber \pder{u''_j\xt}{t}+u_k\xt\pder{u''_j\xt}{x_k}+u''_k\xt\pder{\tu_j\xt}{x_k}-\frac{1}{\bar{\rho}\xt}\pder{R_{jk}\xt}{x_k} =  &\\
\frac{1}{\rho\xt}\pder{\sigma'_{jk}\xt}{x_k}+\lr{\spv\xt-\frac{1}{\bar{\rho}\xt}}\pder{\bar{\sigma}_{jk}\xt}{x_k},& \label{primj2}\\
\nonumber \pder{\rho\xt u''_j\xt}{t}+\pder{\rho\xt u''_j\xt u_k\xt}{x_k}+\rho\xt u''_k\xt\pder{\tu_j\xt}{x_k}-\frac{\rho\xt}{\bar{\rho}\xt}\pder{R_{jk}\xt}{x_k} = & \\
\pder{\sigma'_{jk}\xt}{x_k}+\lr{1-\frac{\rho\xt}{\bar{\rho}\xt}}\pder{\bar{\sigma}_{jk}\xt}{x_k}.\label{consj2}
\end{align}
\end{subequations}
The two-point correlation is defined as,
\begin{equation}
Q_{ij}\lr{x_1,x_2}=\frac{1}{2}\overline{\left[\rho(x_1)+\rho(x_2)\right]u''_i(x_1)u''_j(x_2)}.
\end{equation}
The time derivative of the unaveraged product is,
\begin{align}
&\frac{1}{2}\pder{\lr{\rho(x_1)+\rho(x_2)}u''_i(x_1)u''_j(x_2)}{t} = \frac{1}{2}\pder{\rho(x_2)u''_j(x_2)u''_i(x_1)}{t}+\frac{1}{2}\pder{\rho(x_1)u''_i(x_1)u''_j(x_2)}{t} \nonumber\\
&=\rho(x_2)u''_j(x_2)\underbrace{\pder{u''_i(x_1)}{t}}_\eqref{primi1}+u''_i(x_1)\underbrace{\pder{\rho(x_2)u''_j(x_2)}{t}}_\eqref{consj2}+\rho(x_1)u''_i(x_1)\underbrace{\pder{u''_{j}(x_2)}{t}}_\eqref{primj2}+u''_j(x_2)\underbrace{\pder{\rho(x_1)u''_i(x_1)}{t}}_\eqref{consi1}
\end{align}
Therefore, an exact equation for the two-point correlation is obtained by performing these opeartions on the above equations and averaging.
We note that $x_1$ and $x_2$ are independent and the spatial derivatives at these locations are denoted by subscripts 1 or 2.
Each of these terms is discussed below.\\
\noindent Terms featuring $\tu_k(x_1)$: 
\begin{multline}
\ol{\rho(x_2)u''_j(x_2)\tu_k\xo\pder{u''_i\xo}{x_{k_1}}}+\ol{u''_{j}(x_2)\pder{\rho\xo u''_i\xo\tu_k\xo}{x_{k_1}}} = \\
\ol{\pder{\rho(x_2)u''_j(x_2)u''_i\xo\tu_k\xo}{x_{k_1}}}+\ol{\pder{\rho\xo u''_j(x_2)u''_i\xo\tu_k\xo}{x_{k_1}}}
-\ol{\rho(x_2)u''_i\xo u''_j(x_2)}\pder{\tu_k\xo}{x_{k_1}} = \\
\pder{Q_{ij}(x_1,x_2)\tu_k\xo}{x_{k_1}}-\frac{1}{2}\ol{\rho(x_2)u''_i\xo u''_j(x_2)}\pder{\tu_k\xo}{x_{k_1}}
\end{multline}
The terms with $\tu_k\xt$ also reduce to a similar form and collecting the terms with minus sign from these,
\begin{multline}
\frac{1}{2}\ol{\rho\xo u''_i\xo u''_j}\xt\pder{\tu_k\xo}{x_{k_1}}+\frac{1}{2}\ol{\rho\xo u''_j\xt u''_i\xo}\pder{\tu_k\xt}{x_{k_2}}= \\
\frac{1}{2}\ol{\lrs{\rho\xt-\frac{\rho\xo}{2}+\frac{\rho\xo}{2}}u''_i\xo u''_j}\xt\pder{\tu_k\xo}{x_{k_1}}+ \\
\frac{1}{2}\ol{\lrs{\rho\xo-\frac{\rho\xt}{2}+\frac{\rho\xt}{2}}\rho\xo u''_j\xt u''_i\xo}\pder{\tu_k\xt}{x_{k_2}}= \\
\frac{1}{4}\ol{\lrs{\rho\xo+\rho\xt}u''_i\xo u''_j\xt}\pder{\tu_k\xo}{x_{k_1}}+\frac{1}{4}\ol{\lrs{\rho\xo+\rho\xt}u''_i\xo u''_j\xt}\pder{\tu_k\xt}{x_{k_2}} -\\
\frac{1}{4}\ol{\lrs{\rho\xt-\rho\xo}u''_i\xo u''_j\xt}\pder{\tu_k\xo}{x_{k_1}}+\frac{1}{4}\ol{\lrs{\rho\xo-\rho\xt}u''_i\xo u''_j\xt}\pder{\tu_k\xt}{x_{k_2}}= \\
\frac{1}{2}Q_{ij}(x_1,x_2)\left[\pder{\tu_k\xo}{x_{k_1}}+\pder{\tu_k\xt}{x_{k_2}}\right]+ 
\frac{1}{2}Q_{ij}^{(-)}(x_1,x_2)\left[\pder{\tu_k\xt}{x_{k_2}}-\pder{\tu_k\xo}{x_{k_1}}\right] ,
\end{multline}
where $Q_{ij}^{(-)}\lr{x_1,x_2}=\frac{1}{2}\ol{\lrs{\rho\xo-\rho\xt}u''_i\xo u''_j\xt}$.\\
Terms featuring  $\tu_i(x_1)$ and $\tu_j(x_2)$: \\
\begin{align}
\frac{1}{2}\ol{\rho\xt u''_j\xt u''_k\xo}\pder{\tu_i\xo}{x_{k_1}}+\frac{1}{2}\ol{u''_j\xt\rho\xo u''_k\xo}\pder{\tu_i\xo}{x_{k_1}}=Q_{kj}(x_1,x_2)\pder{\tu_i\xo}{x_{k_1}} \\
\frac{1}{2}\ol{\rho\xo u''_i\xo u''_k\xt}\pder{\tu_j\xt}{x_{k_2}}+\frac{1}{2}\ol{u''_i\xo\rho\xt u''_k\xt}\pder{\tu_j\xt}{x_{k_2}}=Q_{kj}(x_1,x_2)\pder{\tu_j\xt}{x_{k_2}}
\end{align}
Triple correlation terms: \\
\begin{multline}
\frac{1}{2}\ol{\rho(x_2)u''_j(x_2)u''_k(x_1)\pder{u''_i(x_1)}{x_{k_1}}}+\frac{1}{2}\ol{u''_j(x_2)\pder{\rho(x_1)u''_i(x_1)u''_k(x_1)}{x_{k_1}}}= \\
\frac{1}{2}\ol{\pder{\rho(x_2)u''_j(x_2)u''_i(x_1)u''_k(x_1)}{x_{k_1}}}+\frac{1}{2}\ol{\pder{\rho(x_1)u''_i(x_1)u''_j(x_2)u''_k(x_1)}{x_{k_1}}}-\frac{1}{2}\ol{\rho(x_2)u''_j(x_2)u''_i(x_1)\pder{u''_k(x_1)}{x_{k_1}}}=\\
\frac{1}{2}\pder{}{x_{k_1}}\ol{\lrs{\rho\xo+\rho\xt}u''_i\xo u''_j\xt u''_k\xo}-\frac{1}{2}\ol{\rho(x_2)u''_j(x_2)u''_i(x_1)\pder{u''_k(x_1)}{x_{k_1}}} 
\end{multline}
The triple correlations from \eqref{primj2} and \eqref{consj2} has a similar form.
The terms with minus sign can be written as,
\begin{multline}
\frac{1}{2}\ol{\rho(x_2)u''_j(x_2)u''_i(x_1)\pder{u''_k(x_1)}{x_{k_1}}}+\ol{\frac{1}{2}\rho(x_1)u''_j(x_2)u''_i(x_1)\pder{u''_k(x_2)}{x_{k_2}}} = \\
\frac{1}{2}\ol{\lrs{\rho(x_2)+\rho(x_1)+\rho(x_2)-\rho(x_1)}u''_j(x_2)u''_i(x_1)\pder{u''_k(x_1)}{x_{k_1}}} \\
+\frac{1}{2}\ol{\lrs{\rho(x_1)+\rho(x_2)+\rho(x_1)-\rho(x_2)}u''_j(x_2)u''_i(x_1)\pder{u''_k(x_2)}{x_{k_2}}} = \\
\frac{1}{2}\ol{\lrs{\rho(x_1)+\rho(x_2)}u''_i(x_1)u''_j(x_2)\pder{u''_k(x_1)}{x_{k_1}}}-\ol{\frac{1}{2}\lrs{\rho(x_1)-\rho(x_2)}u''_i(x_1)u''_j(x_2)\pder{u''_k(x_1)}{x_{k_1}}}  \\
+\frac{1}{2}\ol{\lrs{\rho(x_1)+\rho(x_2)}u''_j(x_2)u''_i(x_1)\pder{u''_k(x_2)}{x_{k_2}}}+\frac{1}{2}\ol{\lrs{\rho(x_1)-\rho(x_2)}u''_j(x_2)u''_i(x_1)\pder{u''_k(x_2)}{x_{k_2}}} =\\
\frac{1}{2}\lrs{H_{ij}\lr{x_1,x_2;x_2}+H_{ji}\lr{x_2,x_1;x_1}}+\frac{1}{2}\lrs{H_{ij}^-\lr{x_1,x_2;x_2}-H_{ji}^-\lr{x_2,x_1;x_1}},
\end{multline}
where,
\begin{align}
H_{ij}\lr{x_1,x_2;x_2}=1/2\ol{\lrs{\rho\xo+\rho\xt}u''_i\xo u''_j\xt\pder{u''_k\xt}{x_{k_2}}} \\
H_{ij}^-\lr{x_1,x_2;x_2}=1/2\ol{\lrs{\rho\xo-\rho\xt}u''_i\xo u''_j\xt\pder{u''_k\xt}{x_{k_2}}}. 
\end{align}
Alternatively, these can be simplified to,
\begin{multline}
\frac{1}{2}\lrs{H_{ij}\lr{x_1,x_2;x_2}+H_{ji}\lr{x_2,x_1;x_1}}+\frac{1}{2}\lrs{H_{ij}^-\lr{x_1,x_2;x_2}-H_{ji}^-\lr{x_2,x_1;x_1}}=\\
\frac{1}{2}\ol{\rho\xt u''_i\xo u''_j\xt\pder{u''_k\xo}{x_{k_1}}}+\frac{1}{2}\ol{\rho\xo u''_i\xo u''_j\xt\pder{u''_k\xt}{x_{k_2}}}. 
\end{multline}
Fluctuating stress terms:
\begin{equation}
\frac{1}{2}\ol{\rho(x_2)u''_j(x_2)\spv(x_1)\pder{\sigma'_{ik}(x_1)}{x_{k_1}}}+\frac{1}{2}\ol{u''_j(x_2)\pder{\sigma'_{ik}(x_1)}{x_{k_1}}}=
\frac{1}{2}\ol{u''_j(x_2)\lrs{1+\frac{\rho(x_2)}{\rho(x_1)}}\pder{\sigma'_{ik}(x_1)}{x_{k_1}}} =\frac{1}{2}\Psi_{ji}(x_2,x_1)
\end{equation}
Then by noting that $x_2=x_1+r$, the derivatives can be written as (F. Hamba, private communication, April 2018),
\begin{equation}
\pder{u''_j\xt}{x_{k_2}}=\pder{u''_j\lr{x_1+r}}{r_k}=\pder{u''_j\xt}{x_{k_1}}.
\label{hamba}
\end{equation}
\begin{align}
\therefore \Psi_{ji}\lr{x_2,x_1}&=\ol{\lrs{1+\frac{\rho(x_2)}{\rho(x_1)}}\pder{}{x_{k_1}}u''_j\xt\sigma'_{ik}\xo}-\ol{\lrs{1+\frac{\rho(x_2)}{\rho(x_1)}}\sigma'_{ik}\xo\pder{u''_j\xt}{x_{k_2}}} \nonumber \\
&=\ol{\lrs{1+\frac{\rho(x_2)}{\rho(x_1)}}\pder{}{x_{k_1}}u''_j\xt\tau'_{ik}\xo}-\ol{\lrs{1+\frac{\rho(x_2)}{\rho(x_1)}}\tau'_{ik}\xo\pder{u''_j\xt}{x_{k_2}}} \nonumber \\
&-\ol{\lrs{1+\frac{\rho(x_2)}{\rho(x_1)}}\pder{}{x_{i_1}}u''_j\xt p\xo}+\ol{\lrs{1+\frac{\rho(x_2)}{\rho(x_1)}}p\xo\pder{u''_j\xt}{x_{i_2}}}
\end{align}
Coeffecient of mean stress gradient $\bar{\sigma}$:
\begin{multline}
-\frac{1}{2}\ol{\rho\xt u''_j\xt\lr{\spv\xo-\frac{1}{\bar{\rho}\xo}}}-\frac{1}{2}\ol{u''_j\xt\lr{1-\frac{\rho\xo}{\bar{\rho}\xo}}}= \\
-\frac{1}{2}\ol{\lrs{\rho\xo+\rho\xt}u''_j\xt\lr{\spv\xo-\frac{1}{\bar{\rho}\xo}}}  \\
=-\frac{1}{2}\ol{\rho\xt u''_j\xt\spv\xo}-\frac{1}{2}\ol{u''_j\xt}+\frac{1}{2}\frac{\ol{\rho\xo u''_j\xt}}{\bar{\rho}\xo}
\end{multline}
Coefficient of turbulent stress:
\begin{equation}
-\frac{1}{2}\ol{\rho\xt u''_j\xt}/\bar{\rho}\xo-\frac{1}{2}\ol{\rho\xo u''_j\xt}/\bar{\rho}\xo=-\frac{1}{2}\frac{\ol{\rho\xo u''_j\xt}}{\bar{\rho}\xo}
\end{equation}
Then by applying a coordinate transformation,
\begin{align}
\nonumber x_1=\mbx, & x_2=\mbx+\mbr, \\
\nonumber \pder{}{x_{k_1}}=\pder{}{x_k}-\pder{}{r_k}, & \pder{}{x_{k_2}}=\pder{}{r_k},
\end{align}
along with the relations \eqref{hamba}, the exact equation for two-point correlation is obtained (equation 2.9 in the manuscript).
The linear and non-linear transport in the scale space are obtained by this transformation.
\begin{align}
\pder{}{x_{k_1}}Q_{ij}\lr{x_1,x_2}\tu_k\xo+\pder{}{x_{k_2}}Q_{ij}\lr{x_1,x_2}\tu_k\xt=&\pder{}{x_k}Q_{ij}\lr{\mbx,\mbr}\tu_k(\mbx) \nonumber \\
&+\pder{}{r_k}Q_{ij}\lr{\mbx,\mbr}\lrs{\tu_k(\mbx+\mbr)-\tu(\mbx)}
\end{align}
\begin{multline}
\pder{}{x_{k_1}}\frac{1}{2}\ol{\lrs{\rho\xo+\rho\xt}u''_i\xo u''_j\xt u''k\xo}+\pder{}{x_{k_2}}\frac{1}{2}\ol{\lrs{\rho\xo+\rho\xt}u''_i\xo u''_j\xt u''k\xt}= \\
\pder{T_{ijk}(\mbx,\mbr)}{x_k}
+\frac{1}{2}\pder{}{r_k}\ol{\lrs{\rho(\mbx)+\rho(\mbx+\mbr)}u''_i(\mbx)u''_j(\mbx+\mbr)\lrs{u''_k(\mbx+\mbr)-u''_k(\mbx)}}
\end{multline}

The transport equation for the trace of the two-point tensor is,
\begin{flalign}
&\pder{Q_{ii}(\mbx,\mbr)}{t}+\pder{Q_{ii}(\mbx,\mbr)\tilde{u}_k(\mbx)}{x_{k}}= 
\underbrace{-Q_{ki}(\mbx,\mbr)\pder{\tilde{u}_i(\mbx)}{x_{k}}-Q_{ik}(\mbx,\mbr)\pder{\tilde{u}_i(\mbx+\mbr)}{x_{k}}}_{\mathcal{P}_{ii}} 
\nonumber \\
&\underbrace{-\pder{T_{iik}(\mbx,\mbr)}{x_{k}}}_{\mathcal{D}^u_{ii}}+\underbrace{\frac{1}{2}\lrs{\Phi_{ii}^p(\mbx,\mbx+\mbr)+\Phi_{ii}^p(\mbx+\mbr,\mbx)}}_{\mathcal{D}_{ii}^p}+\underbrace{\frac{1}{2}\lrs{\Phi_{ii}^\nu(\mbx,\mbx+\mbr)+\Phi_{ii}^\nu(\mbx+\mbr,\mbx)}}_{\mathcal{D}_{ii}^\nu} \nonumber \\
&+\underbrace{\frac{1}{2}\lrs{\Psi_{ii}^p(\mbx,\mbx+\mbr)+\Psi_{ii}^p(\mbx+\mbr,\mbx)}}_{\Pi_{ii}}+\underbrace{\frac{1}{2}\lrs{\Psi_{ii}^\nu(\mbx,\mbx+\mbr)+\Psi_{ii}^\nu(\mbx+\mbr,\mbx)}}_{\epsilon_{ii}} \nonumber \\
&\underbrace{+\frac{1}{2}Q_{ii}(\mbx,\mbr)\lrs{\pder{\tilde{u}_k(\mbx)}{x_{k}}+\pder{\tilde{u}_k(\mbx+\mbr)}{x_{k}}}}_{\chi_{ii}} 
\underbrace{+\frac{1}{2}Q^{(-)}_{ii}(\mbx,\mbr)\lrs{\pder{\tilde{u}_k(\mbx+\mbr)}{x_{k}}-\pder{\tilde{u}_k(\mbx)}{x_{k}}}}_{\chi_{ii}^-} \nonumber \\
&\underbrace{+\frac{1}{2}\overline{\rho(\mbx+\mbr)u''_i(\mbx)u''_i(\mbx+\mbr)\pder{u''_k(\mbx)}{x_k}}+\frac{1}{2}\overline{\rho(\mbx)u''_i(\mbx)u''_i(\mbx+\mbr)\pder{u''_k(\mbx+\mbr)}{x_k}}}_{\delta_{ii}} \nonumber \\
&\underbrace{+\frac{1}{2}\lr{a_i(\mbx,\mbx+\mbr)\pder{\bar{\sigma}_{ik}(\mbx+\mbr)}{x_k}+a_i(\mbx+\mbr,\mbx)\pder{\bar{\sigma}_{ik}(\mbx+\mbr)}{x_k}}}_{\Sigma_{ii}} 
\underbrace{+\frac{1}{2}\lrs{c_{ii}(\mbx,\mbx+\mbr)+c_{ii}(\mbx+\mbr,\mbx)}}_{\mathcal{R}_{ii}} \nonumber \\
&\underbrace{-\pder{}{r_{k}}\lrs{\frac{1}{2}\overline{\lr{\rho(\mbx)+\rho(\mbx+\mbr)}u_i''(\mbx)u_i''(\mbx+\mbr)(u_k''(\mbx+\mbr)-u_k''(\mbx))} 
+\lr{\tilde{u}_k(\mbx+\mbr)-\tilde{u}_k(\mbx)}Q_{ii}(\mbx,\mbr)}}_{\mathcal{T}_{ii}},
\label{qii}
\end{flalign}

\subsection*{Energy density budget}
Energy density is defined as 
\[E(x,r)=-\frac{1}{2}\pder{}{r}Q_{ii}(x,r)\]
A transport equation for energy density is obtained by doing the operation $-\frac{1}{2}\pder{}{r}$ on equation \ref{qii}.
Differentiation with respect to $r$ commutes with those in space and time. \\
Time derivative:
\begin{equation}
-\frac{1}{2}\pder{}{r}\pder{Q_{ii}(\mbx,\mbr)}{t}=\pder{}{t}\lr{-\frac{1}{2}\pder{}{r}Q_{ii}(\mbx,\mbr)}=\pder{E(\mbx,\mbr)}{t}
\end{equation}
The spatial derivative on the LHS of \eqref{qii} yields,
\begin{eqnarray}
\nonumber -\frac{1}{2}\pder{}{r}\lr{\pder{Q_{ii}(\mbx,\mbr)\tu_k(\mbx)}{x_k}}&=&-\frac{1}{2}\pder{}{r}\lr{\tu_k(\mbx)\pder{Q_{ii}(\mbx,\mbr)}{x_k}+Q_{ii}(\mbx,\mbr)\pder{\tu_k(\mbx)}{x_k}} \\
\nonumber &=&\tu_k(\mbx)\pder{}{x_k}\lr{-\frac{1}{2}\pder{Q_{ii}(\mbx,\mbr)}{r}}+\lr{-\frac{1}{2}\pder{Q_{ii}(\mbx,\mbr)}{r}}\pder{\tu_k(\mbx)}{x_k} \\
&=&\tu_k(\mbx)\pder{E(\mbx,\mbr)}{x_k}+E(\mbx,\mbr)\pder{\tu_k(\mbx)}{x_k},
\label{advection}
\end{eqnarray}
of which the first term is retained on the left and the other is coupled with the mean field dilatation effects on the right.

\noindent The mean shear terms on the RHS give,
\begin{align}
\nonumber\pder{}{r}\lr{Q_{ki}(\mbx,\mbr)\pder{\tu_i(\mbx)}{x_k}+Q_{ik}(\mbx,\mbr)\pder{\tu_i(\mbx+\mbr)}{x_k}}=&\pder{Q_{ki}(\mbx,\mbr)}{r}\pder{\tu_i(\mbx)}{x_k}+\pder{}{r}\lr{Q_{ik}(\mbx,\mbr)\pder{\tu_i(\mbx+\mbr)}{x_k}} \\
\nonumber  &+\pder{Q_{ik}(\mbx,\mbr)}{r}\pder{\tu_i(\mbx)}{x_k}-\pder{Q_{ik}(\mbx,\mbr)}{r}\pder{\tu_i(\mbx)}{x_k} \\
\nonumber =&\pder{Q_{ki}(\mbx,\mbr)}{r}\lr{\pder{\tu_i(\mbx)}{x_k}+\pder{\tu_k(\mbx)}{x_i}}  \\
&+\pder{}{r}\lr{Q_{ik}(\mbx,\mbr)\pder{\lr{\tu_i(\mbx+\mbr)-\tu_i(\mbx)}}{x_k}}
\end{align}
From $\chi_{ii}$,
\begin{align}
\nonumber-\frac{1}{4}\pder{}{r}\lr{Q_{ii}(\mbx,\mbr)\lrs{\pder{\tu_k(\mbx)}{x_k}+\pder{\tu_k(\mbx+\mbr)}{x_k}}}=&-\frac{1}{4}\pder{Q_{ii}(\mbx,\mbr)}{r}\lrs{\pder{\tu_k(\mbx)}{x_k}+\pder{\tu_k(\mbx+\mbr)}{x_k}} \\
\nonumber  &         -\frac{1}{4}Q_{ii}(\mbx,\mbr)\pder{}{r}\lrs{\pder{\tu_k(\mbx)}{x_k}+\pder{\tu_k(\mbx+\mbr)}{x_k}} \\
\nonumber  =&\frac{1}{2}E(\mbx,\mbr)\lrs{\pder{\tu_k(\mbx)}{x_k}+\pder{\tu_k(\mbx+\mbr)}{x_k}}  \\
&-\frac{1}{4}Q_{ii}(\mbx,\mbr)\pder{}{r}\lr{\pder{\tu_k(\mbx+\mbr)}{x_k}},
\end{align}
which when coupled with \eqref{advection} and the derivative of $\chi^-_{ii}$ represents the contribution of mean field dilatation on energy density evolution as,
\begin{align}
\chi_r=&-\frac{1}{4}Q_{ii}(\mbx,\mbr)\pder{}{r}\lr{\pder{\tilde{u}_k(\mbx+\mbr)}{x_k}}+\frac{1}{2}E(\mbx,\mbr)\lrs{\pder{\tilde{u}_k(\mbx+\mbr)}{x_k}-\pder{\tilde{u}_k(\mbx)}{x_k}} \nonumber \\
&-\frac{1}{4}\pder{}{r}\lr{Q_{ii}^{(-)}(\mbx,\mbr)\lrs{\pder{\tilde{u}_k(\mbx+\mbr)}{x_k}-\pder{\tilde{u}_k(\mbx)}{x_k}}}.
\end{align}
The other terms are expressed as the derivatives in $r$ and require no further manipulations.
The final form of the energy density budget is given in the main article.

\bibliographystyle{jfm}

\end{document}